\documentclass[conference,10pt]{IEEEtran}
\usepackage{epsfig,rotating,setspace,latexsym,amsmath,epsf,amssymb,amsfonts,bm,subfigure,epstopdf}
\usepackage{cite,authblk}
\usepackage{bbm}
\usepackage{mathtools,amsthm}
\usepackage{algorithm}
\usepackage[noend]{algpseudocode}

\algrenewcommand\algorithmicforall{\textbf{foreach}}
\algrenewcommand\algorithmicindent{.8em}
\usepackage{comment}
\usepackage{color}
\usepackage{mathtools}

\newtheorem{theorem}{Theorem}
\newtheorem{lemma}{Lemma}

\newenvironment{Proof}[1]{\medskip\par\noindent{\bf Proof:\,}\,#1}{{\mbox{\,$\blacksquare$}\par}}
\IEEEoverridecommandlockouts
\allowdisplaybreaks

\begin{document}

\title{Age of Information in the Presence of an Adversary}

\author{Subhankar Banerjee \qquad Sennur Ulukus\\
\normalsize Department of Electrical and Computer Engineering\\
\normalsize University of Maryland, College Park, MD 20742\\
\normalsize  \emph{sbanerje@umd.edu} \qquad \emph{ulukus@umd.edu}}
	
\maketitle

\begin{abstract}
We consider a communication system where a base station serves $N$ users, one user at a time, over a wireless channel. We consider the timeliness of the communication of each user via the age of information metric. A constrained adversary can block at most a given fraction, $\alpha$, of the time slots over a horizon of $T$ slots, i.e., it can block at most $\alpha T$ slots. We show that an optimum adversary blocks $\alpha T$ \emph{consecutive} time slots of a randomly selected user. The interesting \emph{consecutive property} of the blocked time slots is due to the cumulative nature of the age metric.
\end{abstract}

\section{Introduction}
The \emph{quality of service} offered by any wireless network is generally measured in three dimensions, namely, throughput, packet delay and energy efficiency. These metrics are generally used to optimize the usage of system resources such as energy and bandwidth, rather than to optimize the exact user experience. With the recent developments in internet of things, real-time augmented and virtual reality systems powered by the emerging 5G technology, users' \emph{quality of experience} matters. To incorporate this, a new metric called \emph{age of information} has been introduced in order to keep track of the \emph{freshness} of received information \cite{kaul2012real}; see recent surveys \cite{kosta2017age, SunSurvey, YatesSurvey}. In this paper, we consider scheduling algorithms to minimize the age of information in a wireless network in an \emph{adversarial} setting.

Designing algorithms to minimize the age of information in wireless networks is an active area of research; see for example \cite{wireless-ephremides, wireless-modiano, wireless-ulukus}. Majority of works on scheduling in wireless channels to minimize age of information considers stationary channel models with additive noise, path loss, fading and unintentional multi-user interference. When there is an intentional interferer (e.g., a jammer) the stationarity assumption no longer holds. References \cite{banerjee2020fundamental, bhattacharjee2020competitive} consider a system with a non-stationary environment in the sense of an adversary deciding whether the channel will be ``good'' or ``bad'' for the users in an online manner. The adversary in these works is \emph{unconstrained}, thus, in principle, it can choose all channels as ``bad'', but it does not do so, as these papers focus on the competitive ratio, i.e., how an online algorithm does with respect to an offline algorithm with complete knowledge of the channel sequence; choosing ``bad'' for all channels makes the two algorithms identical.  

References \cite{nguyen2017impact, garnaev2019maintaining} consider a system where an adversary jams the legitimate communication to increase the age of information. In \cite{nguyen2017impact, garnaev2019maintaining}, by jamming, the adversary decreases the signal to noise ratio of the legitimate channel, decreasing the effective communication rate, and hence increasing the communication time for a fixed packet size, and consequently, increasing the age of information. Reference \cite{xiao2018dynamic} considers a system where the adversary blocks a time duration so that during this time duration the legitimate communication cannot take place hence increasing the service time, and consequently, increasing the age of information at the receiver. 

In this paper, we consider an adversary which blocks update packets by jamming them. Different than \cite{garnaev2019maintaining, nguyen2017impact}, which \emph{decrease the effectiveness} of an update packet by elongating its service time via decreasing its signal to noise ratio through jamming it, here we model the adversary to \emph{eliminate} an update packet by jamming it. Different than \cite{xiao2018dynamic}, which considers blocking \emph{durations of time} to disable updates, here we block individual update time slots. Different than \cite{banerjee2020fundamental, bhattacharjee2020competitive}, which consider competitive ratio for an unconstrained adversary, here we consider the age for a constrained adversary. 

In this paper, we consider a \emph{constrained} adversary. The adversary is constrained in that it cannot block all channels all the time. The constraint imposed on the adversary may be understood as a power constraint. Of the total $T$ communication slots, we allow the adversary to block up to $\alpha T$ slots of its choosing, where $\alpha<1$ is a given constant. First, we show that a deterministic scheduling algorithm does poorly, in the sense that the age at the receiver increases proportional to $T$. A deterministic scheduling algorithm does poorly because the adversary can jam the correct slots as the slots utilized are known deterministically. On the other hand, a randomized scheduling  algorithm does better in the sense that even though the age at the receiver increases proportional to $T$, it decreases proportional to $N$, the number of users. Thus, a randomized algorithm with a large number of users can yield a satisfactory age performance. We investigate the performance of a randomized algorithm in depth. We show that an optimum adversary (most harmful adversary) blocks $\alpha T$ consecutive slots of a fixed user. Interestingly, the \emph{consecutive property} of the blocked time slots is specific to the age metric, and is a consequence of the cumulative nature of the age metric. As an aside, for instance, if the metric were the throughput, then which $\alpha T$ slots the adversary blocks would be inconsequential, as the throughput would be proportional to $(1-\alpha)T$ in all cases, while this selection is critical with the age metric.

\section{System Model and Problem Formulation}
\begin{figure}[t]
	\centerline{\includegraphics[width = 0.6 \columnwidth]{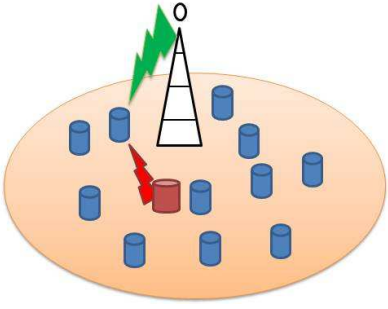}}
	\caption{A base station serves users, and an adversary blocks communication.}
	\label{fig1}
	\vspace*{-0.45cm}
\end{figure}

We consider a system model with a single base station (BS) and $N$ users around the BS; see Fig.~\ref{fig1}. We assume that at any given time slot, the BS can communicate with only one user. The age of the $i$th user at time $t$, is given by $a_{i}(t) = t-u_{i}(t)$, where $u_{i}(t)$ denotes the last time slot at which the $i$th user was served by the BS successfully, before time $t$. Note that the minimum possible age is 1, which is attained if the user is served in the previous time slot. In this work, we have an adversary present in the system which aims to block the communication time slots between the BS and the users. In any time slot, the adversary can block only one user. We consider a \emph{constrained} adversary which cannot block all time slots. Specifically, if the time horizon for the communication is $T$ slots, the adversary can block up to $\alpha$ fraction of the total time horizon $T$. That is, over a total communication duration of $T$ slots, the adversary can block up to $\alpha T$ slots. 

\begin{figure}[t]
    \centerline{\includegraphics[width = 0.79 \columnwidth]{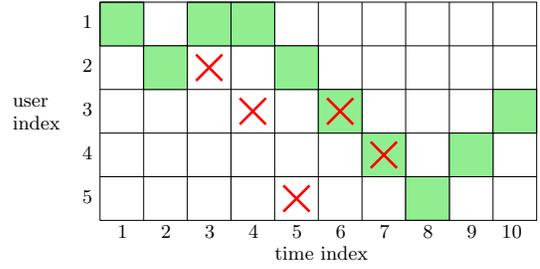}}
    \caption{The base station serves users in green shaded time slots, and the adversary blocks communication in red crossed time slots.}
    \label{fig2}
    \vspace*{-0.4cm}
\end{figure}

We denote the action of the adversary against the $i$th user at time $t$ as $\sigma_{i}(t)$. If the adversary blocks the communication channel for the $i$th user at time $t$, then $\sigma_{i}(t)=0$, and if the adversary does not block the communication channel for the $i$th user at time $t$, then $\sigma_{i}(t)=1$. We define a blocking sequence matrix $\sigma$ of size $N \times T$ whose $(i,t)$ element is $\sigma_{i}(t)$. We define the $i$th row of the matrix $\sigma$ as $\sigma_{i}$, and call it a blocking sequence for user $i$. If a user is successfully served by the BS at a certain time slot, then the age of that user is set to $1$, otherwise it is increased by $1$ with each non-serving time slot. A pictorial representation of the system is shown in Fig.~\ref{fig2}, where for a system with $N=5$ users and $T=10$ time slots, a green shaded $(i,t)$ square shows a time slot $i$ in which the BS serves user $i$, and a red crossed $(i,t)$ square shows a time slot in which the adversary blocks user $i$. In this work, the adversary generates the entire jamming sequence without seeing the action taken by the online user scheduling algorithm. If the user scheduling algorithm is deterministic, the adversary almost surely knows it. 

Formally, age for a user scheduling algorithm $\pi$ (run by the BS) and a blocking matrix $\sigma$ (run by the adversary) is
\begin{align} \label{objective}
\Delta^{\pi,\sigma} = \limsup_{T \to \infty} \frac{1}{T} \sum_{t=1}^{T} \frac{1}{N} \left(\sum_{i=1}^{N} \mathbb{E}\left[a_{i}(t)\right]\right)
\end{align}
where $a_{i}(t)$ is the age of the $i$th user at time $t$. We denote $\Delta_{i}(t)=E[a_{i}(t)]$, the average age of the $i$th user at time $t$. We also denote the overall age of user $i$ as
\begin{align}
\Delta_i=\frac{1}{T} \sum_{t=1}^T \Delta_i(t)
\end{align}
 
The adversary aims to increase the age and the scheduling algorithm aims to reduce the age. We study the interaction between the adversary and the algorithm. Formally, we consider
\begin{align}\label{eq:prob_formu}
      \Delta^{*} = \sup_{\sigma} \inf_{\pi} \quad & \Delta^{\pi,\sigma} \nonumber \\  
       \textrm{s.t.}  \quad & \sum_{i=1}^{N} \sum_{j=1}^{T} 
       (1 - \sigma_i(j)) \leq \alpha T\nonumber\\ 
      & \sum_{i=1}^{N} (1-\sigma_i(j)) \leq 1, \quad j=1,\ldots,T
\end{align}
where $\Delta^{*}$ is the final age. The first constraint is due to the power constraint of the adversary, and the second constraint is due to the fact that at any given time slot the adversary can block at most 1 user. For simplicity of notation, we drop $\pi$ and $\sigma$ from the superscript of $\Delta$ in the rest of the paper.
   
\section{Analysis of Age}
We begin this section with two simple lemmas. Lemma~\ref{lemma:1} below states that any deterministic user scheduling algorithm, which chooses the order in which users are served by the BS deterministically, achieves an age that is lower bounded by a function that increases linearly with $T$ that does not depend on $N$. Lemma~\ref{lemma:2} below, on the other hand, states that if the algorithm chooses the user to be served by the BS uniformly at random, even though the average age is again lower bounded by a term that increases linearly with $T$, the coefficient of the linear increase is inversely proportional with $N$. That is, if the number of users $N$ is large, then the age can decrease to a satisfactory level, despite the presence of an adversary. 

\begin{lemma}\label{lemma:1}
 For any deterministic user scheduling algorithm, the age is lower bounded by $\frac{T \alpha^2}{2}$.
\end{lemma}

\begin{Proof}
Since the algorithm is deterministic, the adversary knows which user is being served in each time slot. Since we are finding a lower bound, we consider that the adversary blocks consecutive $\alpha T$ time slots, blocking the user being served in each time slot. During this blocking time interval, the ages of all users increase, because no one is served. Since we are finding a lower bound, assume that the ages of all users start from 1, i.e., the ages of all users increase from 1 to $\alpha T$ in this interval, and also, assume that the ages in the remaining $(1-\alpha)T$ slots are all 1. Thus, age user $i$ is lower bounded as,
\begin{align}
    \Delta_{i} \geq \frac{1}{T} \left( \sum_{t=1}^{\alpha T} t + \sum_{t=1}^{(1-\alpha) T} 1 \right) \geq \frac{T\alpha^{2}}{2} + (1-\alpha)
\end{align}
Thus, the overall age is lower bounded as
\begin{align}
    \Delta = \frac{1}{N} \sum_{i=1}^{N}\Delta_{i} \geq \frac{T\alpha^{2}}{2} + (1-\alpha) \geq \frac{T\alpha^{2}}{2}
\end{align}
which is linear in $T$.
\end{Proof}
 
\begin{lemma} \label{lemma:2}
For any randomized user scheduling algorithm, the age is lower bounded by $\frac{T\alpha^{2}}{2 N }$.    
\end{lemma}

\begin{Proof}
Since we we are finding a lower bound, we consider that the adversary blocks consecutive $\alpha T$ time slots of a fixed user. Thus, for these $\alpha T$ time slots, one of the users' age increases from 1 to $\alpha T$ assuming that it starts at 1, also that the ages of all users anywhere else is 1, both of which are due to the fact that we are finding a lower bound. The age of the blocked user is lower bounded as 
\begin{align}
    \Delta_{i} \geq \frac{1}{T} \left( \sum_{t=1}^{\alpha T} t + \sum_{t=1}^{(1-\alpha) T} 1 \right) \geq \frac{T\alpha^{2}}{2} + (1-\alpha)
\end{align}
and the ages of all other users are lower bounded as
\begin{align}
    \Delta_{k} \geq \frac{1}{T} \left( \sum_{t=1}^{T} 1 \right) \geq 1
\end{align}
Thus, the overall age is lower bounded as
\begin{align}
\Delta = \frac{1}{N} \sum_{i=1}^{N}\Delta_{i} \geq \frac{1}{N} \left(\frac{T\alpha^{2}}{2} + (1-\alpha) +(N-1) \right) \geq \frac{T\alpha^{2}} {2 N}    
\end{align}
which is linear in $T$, but inverse linear in $N$.
\end{Proof}

The lower bounds in Lemmas~\ref{lemma:1}~and~\ref{lemma:2} hint at the possibility that we may be able to achieve a lower average age for the system by using a randomized scheme, in the presence of an adversary. In the rest of this paper, we will carefully analyze the age performance of a randomized algorithm in the presence of an adversary. First, we will show that, when the algorithm is random, an optimal adversary blocks $\alpha T$ consecutive slots of a randomly chosen user. We find this \emph{consecutive} nature of the optimum blocking strategy of the adversary noteworthy, which is inherent to the age metric. For instance, if the metric was throughput, an optimal adversary would block any randomly chosen $\alpha T$ time slots; whether the blocked time slots are consecutive or not does not have any bearing on the throughput metric, while it is crucial for the age metric. 

To analyze the performance of randomized algorithms in the presence of an adversary, we first note the following facts: If the algorithm does not choose the $i$th user at time $t$, then irrespective of $\sigma_{i}(t)$, $a_{i}(t)$ increases by $1$, i.e., $a_i(t+1)=a_i(t)+1$. The probability of this event is $\frac{N-1}{N}$. If the algorithm chooses the $i$th user at time $t$, then $a_{i}(t+1)$ depends on $\sigma_{i}(t)$. If $\sigma_{i}(t)=0$, i.e., the adversary blocks user $i$, then $a_{i}(t+1)=a_{i}(t)+1$. If $\sigma_{i}(t)=1$, i.e., the adversary does not block user $i$, then $a_{i}(t+1)=1$. Thus, if the algorithm chooses the $i$th user at time $t$, we can write $a_{i}(t+1)$ in a compact form as $a_{i}(t+1)=a_{i}(t)(1-{\sigma}_{i}(t))+1$. The probability of the algorithm choosing user $i$ is $\frac{1}{N}$. Thus, the expected age for user $i$ at time $t+1$ conditioned on the age at time $t$ is
\begin{align}\label{eq:basic}
 \mathbb{E}\left[a_{i}(t+1)|a_{i}(t)\right] = & \frac{N-1}{N} \left(a_{i}(t)+1\right) \nonumber\\ 
 & + \frac{1}{N} \left(a_{i}(t)(1-{\sigma}_i(t))+1\right)   
\end{align}
which simplifies to
\begin{align}\label{eq:7}
 \mathbb{E}[a_{i}(t+1)|a_{i}(t)] = a_{i}(t)\left(1-\frac{{\sigma}_{i}(t)}{N}\right)+ 1
\end{align}
Note that, 
\begin{align}\label{eq:8}
    \mathbb{E}\left[\mathbb{E}\left[a_{i}(t+1)|a_{i}(t)\right]\right] = \Delta_{i}(t+1)
\end{align}
Then, (\ref{eq:7}) and (\ref{eq:8}) give
\begin{align}\label{eq:9}
 \Delta_i(t+1) = \Delta_i(t)\left(1-\frac{{\sigma}_{i}(t)}{N}\right)+ 1
\end{align}
showing how age of user $i$, $\Delta_i(t)$, evolves from time $t$ to $t+1$ as a function of $\sigma_i(t)$.  
Writing one more recursion, we obtain
\begin{align}\label{eq:10}
 \Delta_i(t+1) = & \Delta_i(t-1)\left(1-\frac{{\sigma}_{i}(t-1)}{N}\right)\left(1-\frac{{\sigma}_{i}(t)}{N}\right) \nonumber\\
 &+ \left(1-\frac{{\sigma}_{i}(t)}{N}\right)+1
\end{align}
Proceeding similarly all the way back to time $t=1$ gives
\begin{align}\label{eq:11}
 \Delta_i(t\!+\!1)\!=\!\Delta_i(1) \prod_{j=1}^t \left(1\!-\!\frac{{\sigma}_{i}(j)}{N}\right) 
 \!+\! \sum_{\ell=2}^{t} \prod_{j=\ell}^t \left(1\!-\!\frac{{\sigma}_{i}(j)}{N}\right)\!+\!1
\end{align}

\begin{figure}
	\centerline{\includegraphics[width=0.8\columnwidth]{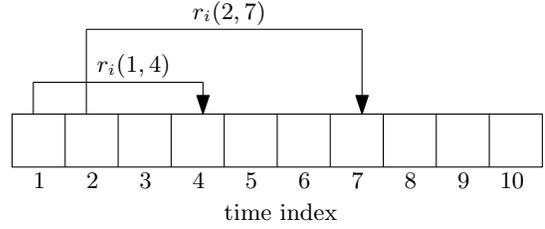}}
	\caption{Example trains for user $i$, $r_i(1,4)$ and $r_i(2,7)$.}
	\label{fig:eq9}
	\vspace*{-0.4cm}
\end{figure}

Next, we define a \emph{train} $r_{i}(k,\ell)$ for user $i$ as a series of time slots starting at slot $k$ and going up to slot $\ell$. There are $(\ell-k+1)$ elements in the train $r_i(k,\ell)$, and each train can be uniquely identified with its starting time and end time. As an example, Fig.~\ref{fig:eq9} shows trains $r_i(1,4)$ and $r_i(2,7)$ with arrows, where the starting time of the train is the starting point of the arrow and the end time is the end point of the arrow. Next, we define the \emph{value of a train} $\Gamma_i(k,\ell)$ as the multiplication of the terms $\left(1-\frac{{\sigma}_{i}(j)}{N}\right)$ for time slots $j$ in the train $r_i(k,\ell)$,  
\begin{align} \label{def:train}
  \Gamma_i(k,\ell)=\prod_{j=k}^\ell \left(1-\frac{{\sigma}_{i}(j)}{N}\right)
\end{align}
Note that each multiplicative term in the train is either $1$ if $\sigma_{i}(j)=0$ (i.e., adversary does not jam slot $j$ for user $i$) or $\frac{N-1}{N}$ if $\sigma_i(j)=1$ (i.e., adversary jams slot $j$ for user $i$).

We can write $\Delta_i(t+1)$ in (\ref{eq:11}) using $\Gamma_i(k,\ell)$ in (\ref{def:train}) as,
\begin{align}\label{eq:12}
 \Delta_i(t+1)=\Delta_i(1) \Gamma_i(1,t)
 + \sum_{\ell=2}^{t} \Gamma_i(\ell,t)+1
\end{align}
Further, noting that $\Delta_i(1)=1$, (\ref{eq:12}) becomes
\begin{align}\label{eq:13}
 \Delta_i(t+1)= \sum_{\ell=1}^{t} \Gamma_i(\ell,t)+1
\end{align}
That is, the average age at time $(t+1)$ is the sum of the values of $t$ trains each starting from $\ell$ where $\ell=1,\ldots,t$ and ending at $t$, and further, the value of each train is a multiplication of a sequence of values $\frac{N-1}{N}$ or $1$ at each time instance depending on whether the adversary jams or not jams, respectively. 

As noted above, from (\ref{eq:13}), the average age in time slot $(t+1)$, $\Delta_i(t+1)$, depends on the adversarial actions from time slot $1$ to time slot $t$. To simplify our upcoming proofs, we assume that (\ref{eq:13}) is the expression for the average age in time slot $t$, i.e., $\Delta_i(t)$, and not time slot $(t+1)$, i.e., $\Delta_i(t+1)$. That is, we shift the age of any time slot to its previous time slot, without loss of optimality. This new assumption does not change the original problem (\ref{eq:prob_formu}), as the age at time $1$ is $1$, which is independent of any adversarial actions.

If a sequence $\sigma_{i}(t)$, $t\in\{1,\ldots,T\}$, has consecutive zeros in an interval, then we call that ${\sigma}_{i}$ a \emph{consecutive blocking sequence} (CBS). The number of ones to the left of the zeros of a CBS ${\sigma}_{i}$ is denoted as $L({\sigma}_{i})$ and the number of ones to the right side of the zeros of a CBS ${\sigma}_{i}$ is denoted as $R({\sigma}_{i})$. 

In the following lemma, we prove that for two CBSs with equal number of zeros, if the number of ones to the left of the first CBS is equal to the number of ones to the right of the second CBS, then these CBSs yield the same  average age. 

\begin{lemma}\label{lemma:3}
Let $\bar{\sigma}_{i}$ and $\Tilde{\sigma}_{i}$ be two CBSs with equal number of zeros, and $L(\bar{\sigma}_{i})=R(\Tilde{\sigma}_{i})$, $L(\Tilde{\sigma}_{i})=R(\bar{\sigma}_{i})$. Then, $\bar{\Delta}_{i}  = \Tilde{\Delta}_{i}$.
\end{lemma}

\begin{Proof}
We first note that
\begin{align}\label{eq:14}
 \sum_{t=1}^T  \sum_{\ell=1}^{t} \Gamma_i(\ell,t)
 = \sum_{t=1}^T  \sum_{\ell=t}^{T} \Gamma_i(t, \ell)
\end{align}
The equality in (\ref{eq:14}) can be interpreted as follows: From (\ref{eq:13}), the average age at time slot $t$ is a sum of $t$ trains, where the starting points of these trains vary from $1$ to $t$ and the end points of the trains is fixed at $t$. Now, (\ref{eq:14}) shows that the same sum can be written as a sum of trains whose starting points are fixed at $t$ and the end points vary from $T, T-1,\ldots,t$. Exchanging the starting/end points of these latter set of trains, they may be viewed as starting at $T, T-1,\ldots,t$ and ending at $t$ moving backwards. Thus, the time can be thought of running from $1$ to $T$, and equivalently running from $T$ to $1$. Since the second CBS is equivalent to the first CBS when the time runs from $T$ to $1$, both CBSs yield the same age for the user. 
\end{Proof}

Next, in the following lemma, we prove that if we move a CBS to right or left in such a way to increase the minimum of the number of ones on the left or the right, in other words, if we \emph{center} the CBS better, then the average age increases. That is, an adversary which \emph{centers} the CBS better over the time horizon causes more harm to the system.

\begin{lemma}\label{lemma:4}
Let CBS $\bar{\sigma}_{i}$ yield average age $\bar{\Delta}_{i}$. Let us create a new CBS  $\Tilde{\sigma}_{i}$ by applying a circular shift on $\bar{\sigma}_{i}$ to the right (or left) by one time slot. Let this new CBS yield average age $\Tilde{\Delta}_{i}$. If $\min(L(\bar{\sigma}_{i}),R(\bar{\sigma}_{i}))\leq\min(L(\Tilde{\sigma}_{i}),R(\Tilde{\sigma}_{i}))$, then $\bar{\Delta}_{i} \leq \Tilde{\Delta}_{i}$.
\end{lemma}

\begin{figure}
  \centerline{\includegraphics[width=0.7\columnwidth]{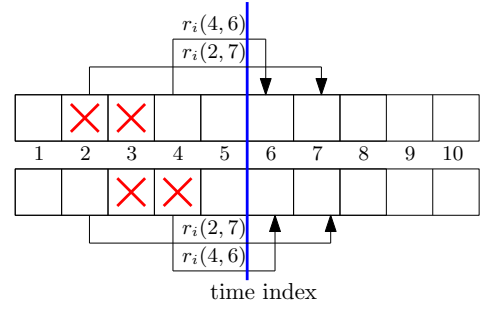}}
  \caption{Pictorial representation of the proof of Lemma~\ref{lemma:4}. Top sequence denotes CBS $\bar{\sigma}_i$ and the bottom sequence denotes CBS $\tilde{\sigma}_i$ which is shifted to the right by one. Till the vertical line both CBSs generate the same average age. After the vertical line, the average age generated by the bottom CBS is larger.}
  \label{fig:4}
  \vspace*{-0.4cm}	
\end{figure}

\begin{Proof}
Without loss of generality, consider that we circularly shift $\bar{\sigma}_{i}$ to the right by one time slot and the condition $\min(L(\bar{\sigma}_{i}),R(\bar{\sigma}_{i}))\leq\min(L(\Tilde{\sigma}_{i}),R(\Tilde{\sigma}_{i}))$ holds true. This implies that $R(\bar{\sigma}_{i})\geq L(\bar{\sigma}_{i})+1$, which further implies $R(\Tilde{\sigma}_{i})\geq L(\bar{\sigma}_{i})$. Thus, we can choose a time slot $t_{1}\in\{0,1,\ldots,T\}$ such that the whole zero portions of both of the CBSs lie in the interval $[0,t_{1}]$, and the number of $1$s to the right of $\Tilde{\sigma}_{i}$ till time $t_{1}$ is the same as $L(\bar{\sigma}_{i})$. From Lemma~\ref{lemma:3}, it follows that $\sum_{t=1}^{t_{1}}\bar{\Delta}_{i}(t)  = \sum_{t=1}^{t_{1}} \Tilde{\Delta}_{i}(t)$. A pictorial representation of this case is given in Fig.~\ref{fig:4}, where till the vertical line, which represents $t_{1}$, both sequences give the same average age. Now, for each time slot $t_{1}\leq t \leq T$, the train $r_i(j,t)$, $t_{1}\leq j \leq t$, corresponding to $\Tilde{\sigma}_{i}$ has equal or larger number of $0$s compared to the train with the same starting and end times corresponding to ${\bar{\sigma}_{i}}$. Note that, for two equal-length trains, the train with more $0$s (more blockage) provides more average age, as depending on $\sigma_{i}(t)$, the multiplier in $\Gamma_i(j,t)$ at time $\tau$, where $j\leq \tau\leq t$, can take value either $1$ (if $\sigma_i(\tau)=0$) or $\frac{N-1}{N}$ (if $\sigma_i(\tau)=1$). This is shown with example trains $r_i(4,6)$ and $r_i(2,7)$ in Fig.~\ref{fig:4} for CBSs $\bar{\sigma}_i$ (top curve) and  $\tilde{\sigma}_i$ (bottom curve). Note that the number of $0$s or $1$s in the train $r_i(2,7)$ did not change, thus $\Gamma_i(2,7)$ is same for both the CBSs, while the number of zeros in the train $r_i(4,6)$ increased, thus increasing $\Gamma_i(4,6)$ going from $\bar{\sigma}_i$ to $\tilde{\sigma}_i$. These observations imply that $\sum_{t=t_{1}+1}^{T}\bar{\Delta}_{i}(t) \leq \sum_{t=t_{1}+1}^{T} \Tilde{\Delta}_{i}(t)$. Hence, $\bar{\Delta}_{i} \leq \Tilde{\Delta}_{i}$.
\end{Proof}

For a general blocking sequence, adversary may have multiple blocks of consecutive zeros, that is, it may have blocks of ones in between blocks of zeros. In the following lemma, we prove that, instead of having multiple disconnected blocks of zeros, an optimum adversary (an adversary that gives the most harm) should have a single connected block of zeros. 

\begin{lemma}\label{lemma:5}
Among all the blocking sequences which has the same number of zeros, a CBS ${\sigma}_{i}$ with either $L({\sigma}_{i})=R({\sigma}_{i})$ or $|L({\sigma}_{i})-R({\sigma}_{i})|\leq 1$ provides the maximum average age. 
\end{lemma}

\begin{Proof}
Without loss of generality, consider a blocking sequence, $\hat{\sigma}_{i}$ which has two blocks of zeros separated by a block of ones, as illustrated in Fig.~\ref{fig6}. Thus, $\hat{\sigma}_i$ is a blocking sequence which consists of two CBSs, $\bar{\sigma}_{i}$ and $\tilde{\sigma}_{i}$, as depicted in Fig.~\ref{fig6}. Let $\bar{l}$ and $\tilde{l}$ be the lengths of the blocks of zeros of $\bar{\sigma}_{i}$ and $\tilde{\sigma}_{i}$, respectively. We call the block of zeros corresponding to $\tilde{\sigma}_{i}$ in $\hat{\sigma}_{i}$, and the block of zeros corresponding to $\bar{\sigma}_{i}$ in $\hat{\sigma}_{i}$, the right block and the left block of $\hat{\sigma}_{i}$, respectively. Without loss of generality, we assume that $T-(\bar{l}+\tilde{l})$ is an even number, i.e., a CBS ${\sigma}_{i}$ is feasible with $(\bar{l}+\tilde{l})$ length of zeros and $L({\sigma}_{i})=R({\sigma}_{i})$. Note that, if $T-(\bar{l}+\tilde{l})$ is an odd number, then the other condition holds true, namely, $|L({\sigma}_{i})-R({\sigma}_{i})|\leq 1$. To prove this lemma, we prove two possible cases. 

{\it In the first case,} we assume that $L(\tilde{\sigma}_{i})-\bar{l}$ is strictly greater than $R(\tilde{\sigma}_{i})$. Now, we create another blocking sequence, $\check{\sigma}_{i}$, by applying a left circular shift by one time slot, only to the $\tilde{\sigma}_{i}$ part of $\hat{\sigma}_{i}$, which is illustrated in Fig.~\ref{fig6}. Now we show that, $\sum_{t=1}^{T} \hat{\Delta}_{i}(t) \leq \sum_{t=1}^{T} \check{\Delta}_{i}(t)$. Note that, we can think of $\check{\sigma}_{i}$, as a blocking sequence consisting of two CBSs, namely, $\bar{\sigma}_{i}$ and the left circular shifted version of $\tilde{\sigma}_{i}$, which we call $\tilde{\tilde{\sigma}}_{i}$. We define $t_{2}$ as the time slot at which the block of zeros for $\tilde{\tilde{\sigma}}_{i}$ starts. For any $t'$, $0\leq t' < t_{2}$ $\hat{\Delta}_{i}(t') = \check{\Delta}_{i}(t')$. This is true because if we consider a train $r_{i}(j,k)$, $0\leq j \leq k$, $j\leq k <t_{2}$ corresponding to $\hat{\sigma}_{i}$, and another train with the same starting and end points corresponding to $\check{\sigma}_{i}$, they have the same value $\Gamma_i(j,k)$. From the time slot $t_{2}$ onwards the average age for the two sequences differ. Note that for all the time slots $t'$, $t_{2}\leq t' < t_{2}+\tilde{l}$, $\hat{\Delta}_{i}(t') \leq \check{\Delta}_{i}(t')$, and for all the time slots $t'$, $t_{2}+\tilde{l}\leq t' \leq T$,  $\hat{\Delta}_{i}(t') \geq \check{\Delta}_{i}(t')$. The reason for this is explained in the next paragraph. 

\begin{figure}[t]
    \centerline{\includegraphics[width = 0.7\columnwidth]{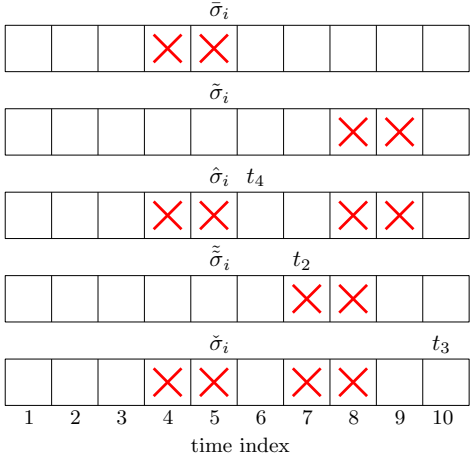}}
    \vspace*{-0.1cm}
    \caption{Pictorial representation of the proof of Lemma~\ref{lemma:5}. The original blocking sequence $\hat{\sigma}_i$ (third row) is composed of two sequences: left part $\bar{\sigma}_i$ (first row) and right part $\tilde{\sigma}_i$ (second row). The modified blocking sequence $\check{\sigma}_i$ (bottom row) is obtained by shifting the right part $\tilde{\sigma}_i$ to left by one time slot which yields $\tilde{\tilde{\sigma}}_i$ (fourth row). $\check{\sigma}_i$ is obtained by bringing two parts closer together, and yields a larger age than $\hat{\sigma}_i$. Thus, $\check{\sigma}_i$ is a better blocking sequence for the advisory than $\hat{\sigma}_i$. In this example, $\bar{l}$ and $\tilde{l}$ are equal to $2$, and  $l_{4}=2$.}
    \label{fig6}
    \vspace*{-0.5cm}
\end{figure}

Consider a time slot $t_{3}\geq t_{2}+\tilde{l}$. Note that a train corresponding to $\hat{\sigma}_{i}$ with starting point $j$, $t_{2}<j\leq t_{2}+\tilde{l}$ and ending point $t_{3}$ has always one less $\frac{N-1}{N}$ in the $(1,\frac{N-1}{N})$ pattern compared to a train corresponding to $\check{\sigma}_{i}$ with the same starting and end points. These trains are the only reasons for the fact that $\hat{\Delta}_i(t') \geq \check{\Delta}_i(t')$, where $t_{2}+\tilde{l}\leq t' \leq T$. The number of such trains for a fixed $t_{3}$ is $\tilde{l}$. For a fixed $t_{3}$, we create a set $\mathcal{T}_{\hat{\sigma}_{i},t_{3}}$ with the trains of $\hat{\sigma}_{i}$, where the starting point of those trains is $j$, $t_{2}<j\leq t_{2}+\tilde{l}$  and the end point is $t_{3}$. Let the number of ones till the time slot $t_{3}$ to the right of $\tilde{\tilde{\sigma}}_{i}$ be $l_{4}$. Due to the assumption of this case, we can always find a time slot $t_{4}$ to the left of $\tilde{\sigma}_{i}$ such that in the interval $[t_{4},t_{2}]$ we get exactly $l_{4}$ ones for $\hat{\sigma}_{i}$. Note that, a train with the starting point $t_{4}$ and the end point $j$, $t_{2}\leq j < t_{2}+\tilde{l}$ corresponding to $\check{\sigma}_{i}$ has always one less $\frac{N-1}{N}$ in the $(1,\frac{N-1}{N})$ pattern compared to a train corresponding to $\hat{\sigma}_{i}$ with the same starting and end points. The number of these trains for a fixed $t_{4}$ is again $\tilde{l}$. We create a set $\mathcal{T}_{\check{\sigma}_{i},t_{4}}$ with the trains of $\check{\sigma}_{i}$ where the starting point of those trains is $t_{4}$ and the end point of those trains is $j$, $t_{2}\leq j < t_{2}+\tilde{l}$. For each $t_{3}$ we can always find a $t_{4}$. For a fixed $t_{3}$ and $t_{4}$, for all the elements of $\mathcal{T}_{\check{\sigma}_{i},t_{3}}$, we can always find an element from $\mathcal{T}_{\hat{\sigma}_{i},t_{4}}$ which has the same number of $\frac{N-1}{N}$.  Thus, we conclude that the total average age corresponding to $\check{\sigma}_{i}$ is greater than or equal to the average age corresponding to $\hat{\sigma}_{i}$.

Now, we can move the right block of $\check{\sigma}_{i}$ further left till the assumption of this case breaks. During this process, it may happen that the right block gets connected to the left block, thus the resultant blocking sequence becomes a CBS. In this case, from Lemma~\ref{lemma:4}, we know that among all the CBSs, a CBS $\sigma_{i}$ with $R(\sigma_{i}) = L(\sigma_{i})$ achieves the maximum average age. If this does not happen (i.e., the right and left blocks do not get connected while satisfying the condition of case 1), then we continue to shift the right block of $\check{\sigma}_{i}$ till it violates the assumption of this case and we call that blocking sequence as $\hat{\hat{\sigma}}_{i}$. When it happens we are certain that the number of ones to the right of the left block of $\hat{\hat{\sigma}}_{i}$ is strictly greater than $L(\bar{\sigma}_{i})$. Now, we circularly shift the left block (i.e., $\bar{\sigma}_{i}$ part) of $\hat{\hat{\sigma}}_{i}$ to the right by one time slot. With similar arguments, we can show that the new sequence achieves higher average age. We shift the left block of $\hat{\hat{\sigma}}_{i}$ to the right till it gets connected to the right block and becomes a CBS with the same number of ones to its right and to its left. From Lemma~\ref{lemma:4}, we know that this particular CBS achieves the maximum average age.

{\it In the second case,} we consider that $L(\tilde{\sigma}_{i})-\bar{l}$ is less than $R(\tilde{\sigma}_{i})$. We circularly shift only the left block of $\hat{\sigma}_{i}$ to the right by one time slot. With similar arguments to the first case above, we can show that this new blocking sequence achieves a higher average age. We can do this right shift of the left block till it gets connected to the right block and becomes a CBS, and from Lemma~\ref{lemma:4} we know that among all the CBSs, a CBS $\sigma_{i}$ with $R(\sigma_{i}) = L(\sigma_{i})$ achieves the maximum average age. These two cases prove the statement of the lemma.
\end{Proof}

Lemma~\ref{lemma:5} implies that for a particular user $i$ if the adversary blocks $\alpha_{1} T$ slots, where $\alpha_{1}\leq\alpha$, then the adversary chooses a CBS as the blocking sequence. Next, we will prove that, if the adversary blocks only a single user with a CBS of length $\alpha T$ slots and does not block any other user, then this will give the maximum average age for the system. We state this main result in Theorem~\ref{th:1} below. Before that, we state a technical lemma, which will be used in the proof of Theorem~\ref{th:1}.

\begin{lemma}\label{lemma:6}
Consider $0\leq\beta\leq 1$, and $a,b,c$ that are integers. Then, $\beta^{a-b} - \beta^{a} \leq   \beta^{c-b} - \beta^{c} $ if $c<a$. 
\end{lemma}

\begin{Proof} 
We note
\begin{align}
\beta^{c-b} \!-\! \beta^{c} = \beta^{c}(\beta^{-b} \!- \!1) 
\geq \beta^{a} (\beta^{-b} \!-\!1) =  \beta^{a-b} \!-\! \beta^{a}
\end{align}
proving the desired result.
\end{Proof}

\begin{theorem}\label{th:1}
For the proposed randomized algorithm, the optimal blocking sequence for the adversary is to block consecutive time slots of any particular user. 
\end{theorem}

\begin{Proof}
For simplicity, we prove this in the two-user case. Let us consider the blocking matrix $\bar{\sigma}$, where the blocking length for $\bar{\sigma}_{1}$ is $\alpha_{1} T$ and the blocking length for $\bar{\sigma}_{2}$ is $\alpha_{2} T$, with $\alpha_{1} +\alpha_{2} = \alpha$. Without loss of generality, consider that the adversary blocks the first user consecutively from time slot $t_{1}$ to time slot $t_{2}$ and again consecutively from time slot $t_{3}$ to time slot $t_{4}$. Thus, $t_{2} + t_{4} - t_{1} - t_{3} + 2 = \alpha_{1} T$. Similarly, consider that the adversary blocks the second user consecutively from time slot $t_{1}'$ to time slot $t_{2}'$ and again consecutively from time slot $t_{3}'$ to time slot $t_{4}'$. Thus, $t_{2}' + t_{4}' - t_{1}' - t_{3}' +2  = \alpha_{2} T$. Without, loss of generality, we can assume that $t_{1}'<t_{2}'<t_{1}<t_{2}<t_{3}'<t_{4}'<t_{3}<t_{4}$. This is illustrated in Fig.~\ref{fig7}. 

Now, consider another blocking matrix $\tilde{\sigma}$, where the blocking length for $\tilde{\sigma}_{1}$ is $\alpha T - (t_{4}' - t_{3}' +1)$ and the blocking length for $\tilde{\sigma}_{2}$ is $(t_{4}' - t_{3}' +1 )$. For this blocking matrix, the adversary blocks the first user consecutively from time slot $t_{1}'$ to time slot $t_{2}'$ and again consecutively from time slot $t_{1}$ to time slot $t_{2}$, and finally, consecutively from time slot $t_{3}$ to time slot $t_{4}$. And, the adversary blocks the second user consecutively from time slot $t_{3}'$ to time slot $t_{4}'$ as illustrated in Fig.~\ref{fig7}. We assume that for time $t$, $\bar{\sigma}$ yields average age $\bar{\Delta}(t)$ and $\tilde{\sigma}$ yields average age $\tilde{\Delta}(t)$. Now, we prove that $\bar{\Delta}(T) \leq \tilde{\Delta}(t)$.

Consider a time slot $\bar{t}$, where $\bar{t}<t_{1}$, the train $r_{1}(j,\bar{t})$ corresponding to $\bar{\sigma}_{1}$ has the same $(1,\frac{N-1}{N})$ pattern as the train $r_{2}(j,\bar{t})$ corresponding to $\tilde{\sigma}_{2}$, where $1\leq j\leq\bar{t}$. Similarly, the train $r_{2}(j,\bar{t})$ corresponding to $\bar{\sigma}_{2}$ has the same $(1,\frac{N-1}{N})$ pattern as the train $r_{1}(j,\bar{t})$ corresponding to $\tilde{\sigma}_{1}$. Thus, we conclude that $\bar{\Delta}(\bar{t}) = \tilde{\Delta}(\bar{t})$.

\begin{figure}[t]
    \centerline{\includegraphics[width = 0.7\columnwidth]{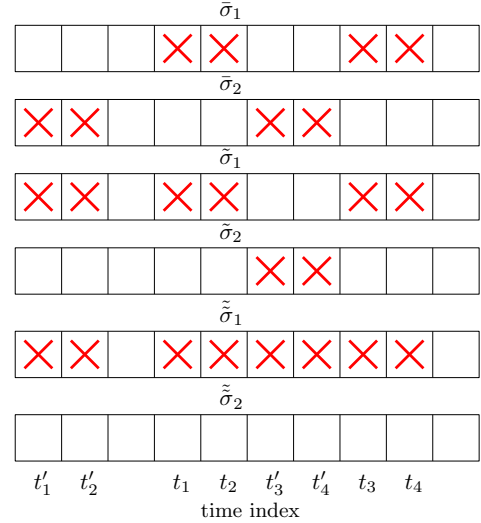}}
    \vspace*{-0.2cm}
    \caption{Pictorial representation of the proof of Theorem~\ref{th:1}. In this figure, $\bar{\sigma}_{1}$ and $\bar{\sigma}_{2}$ constitute $\bar{\sigma}$; $\tilde{\sigma}_{1}$ and $\tilde{\sigma}_{2}$ constitute $\tilde{\sigma}$; and $\tilde{\tilde{\sigma}}_{1}$ and $\tilde{\tilde{\sigma}}_{2}$ constitute $\tilde{\tilde{\sigma}}$. Till time slot $(t_{1} - 1)$, $\bar{\sigma}$ and $\tilde{\sigma}$ yield the same average age. From time slot $t_{1}$ onwards, the average age for the first user corresponding to $\tilde{\sigma}_{1}$ is higher than the average age for the first user corresponding to $\bar{\sigma}_{1}$, and the average age for the second user corresponding to $\bar{\sigma}_{2}$ is higher than the average age for the second user corresponding to $\tilde{\sigma}_{2}$. Thus, if the adversary moves from $\bar{\sigma}$ to $\tilde{\sigma}$, the average age for the first user increases and the average age for the second user decreases. Theorem~\ref{th:1} shows that this increment is more than the decrement. Thus, going from $\bar{\sigma}$ to $\tilde{\sigma}$ then to $\tilde{\tilde{\sigma}}$ increases the age.}
    \label{fig7}
    \vspace*{-0.4cm}
\end{figure}

From the time slot $t_{1}$, the average age for the two sequences differ. Consider a time slot $\bar{\bar{t}} \geq t_{1}$, the value of the train $r_{1}(j,\bar{\bar{t}})$, $\Gamma_{1}(j,\bar{\bar{t}})$ corresponding to $\tilde{\sigma}_{1}$ is $(1- \frac{1}{N})^{\bar{\bar{t}} - (x+ t_{2}' - t_{1}' + 1)} $  and similarly $\Gamma_{1}(j,\bar{\bar{t}})$ corresponding to $\bar{\sigma}_{1}$ is $ (1-\frac{1}{N})^{\bar{\bar{t}} - x} $, where $x$ is an appropriate constant depending on the time slots $\bar{\bar{t}}$ and $j$, where $1 \leq j \leq \bar{\bar{t}} $. For example, in Fig.~\ref{fig7} if  $\bar{\bar{t}}=5$ and $j=1$, then $x=2$. The value of the train $r_{2}(j,\bar{\bar{t}})$, $\Gamma_{2}(j,\bar{\bar{t}})$ corresponding to $\bar{\sigma}_{2}$ is $(1 - \frac{1} {N})^{\bar{\bar{t}} - y - (t_{2}' - t_{1}' +1)}$ and similarly, $\Gamma_{2}(j,\bar{\bar{t}})$ corresponding to $\tilde{\sigma}_{2}$ is $(1 - \frac{1} {N})^{\bar{\bar{t}} - y}$. For example in Fig.~\ref{fig7}, $j=1$ and $\bar{\bar{t}} = 5$, $y=0$. Note that for a fixed $j$ and fixed $\bar{\bar{t}}$, where $\bar{\bar{t}}\geq t_{1}$ and $j\leq\bar{\bar{t}}$, $y<x$. Thus, for a fixed $j$ and $\bar{\bar{t}}$, if the adversary chooses $\tilde{\sigma}$ over $\bar{\sigma}$ as the blocking matrix then $\Gamma_{1}(j,\bar{\bar{t}})$ increases by an amount of $ (1- \frac{1}{N})^{\bar{\bar{t}} - (x+ t_{2}' - t_{1}' + 1)} - (1-\frac{1}{N})^{\bar{\bar{t}} - x}$ and $\Gamma_{2}(j,\bar{\bar{t}})$ decreases by an amount of $(1 - \frac{1} {N})^{\bar{\bar{t}} - y - (t_{2}' - t_{1}' + 1)} -  (1 - \frac{1} {N})^{\bar{\bar{t}} - y}$. As $y<x$, from Lemma~\ref{lemma:6}, we observe that the increment of $\Gamma_{1}(j,\bar{\bar{t}})$ is more than the decrement. Thus, $\bar{\Delta}(T) \leq \tilde{\Delta}(T)$.
 
Now, we create another blocking matrix $\tilde{\tilde{\sigma}}$ where the adversary blocks the first user consecutively from time slot $t_{1}'$ to time slot $t_{2}'$, time slot $t_{1}$ to $t_{2}$, time slot $t_{3}'$ to $t_{4}'$, and finally, from time slot $t_{3}$ to $t_{4}$ and does not block the second user. With similar arguments, we can show that $\tilde{\tilde{\sigma}}$ yields a higher average age than $\tilde{\sigma}$. Finally, using Lemma~\ref{lemma:5}, we conclude that if the adversary blocks consecutive time slots for a single user, it provides the maximum average age.
\end{Proof}

Therefore, the optimum action for the adversary is to: 1) use its entire blocking power of $\alpha T$ slots, 2) block one user only, 3) block that user in consecutive $\alpha T$ slots, and 4) move the blocking sequence to the center of the time horizon $T$ as much as possible. Without loss of generality, let us assume that the adversary blocks user $1$. Now, we develop an upper bound for our proposed algorithm.

\begin{theorem}\label{th:2}
An upper bound on the average age with the proposed algorithm is $\frac{T+1}{2N} +(N-1)$.
\end{theorem}

\begin{Proof}
An upper bound for the age of the first (blocked) user is  $\frac{T(T+1)}{2}$. For other (unblocked) users, we consider that the whole horizon $T$ is divided into several intervals. We construct each interval with consecutive transmitted slots for a user followed by consecutive non-transmitted slots. Note that this is a renewal process. For any such interval, let us assume that the length of the transmitted slots is $\tau_{tr}$ and the length of the non-transmitted slots is $\tau_{ntr}$  such that $\tau=\tau_{tr}+\tau_{ntr}$. Then, 
\begin{align} 
    \sum_{\ell=1}^{\tau}\Delta_{i}(\ell) =& \sum_{\ell=1}^{\tau_{tr}}1 +\sum_{\ell=1}^{\tau_{ntr}} (1+\ell) 
    = \tau_{tr} + \frac{\tau_{ntr}^{2}}{2} + \frac{3 \tau_{ntr}}{2} \label{eq:eq10}
\end{align}
The algorithm chooses user $i$ with probability $q=\frac{1}{N}$, thus, $\tau_{tr}$ and $\tau_{ntr}$ have the following geometric distributions: 
$\mathbb{P}(\tau_{tr}=k) = q^{k-1}(1-q)$, and 
$\mathbb{P}(\tau_{ntr}=k) = q(1-q)^{k-1}$, for $k\geq 1$. 

From renewal reward theorem, we know that 
\begin{align}
\lim_{T\to \infty}\mathbb{E}\left[\frac{\sum_{t=1}^{T}\Delta_{i}(t)}{T}\right] = \frac{\mathbb{E}\left[C(\tau)\right]}{\mathbb{E}\left[\tau\right]} \label{dummy1}
\end{align}
Now from (\ref{eq:eq10}), we have
\begin{align}
    \mathbb{E}\left[C(\tau)\right] = \frac{1}{(1-q)} + \frac{3}{2q} + \frac{2-q}{q^{2}} = \frac{1}{q^{2}(1-q)}
    \label{dummy2}
\end{align}
In addition, we have
\begin{align}
    \mathbb{E}[\tau] = \mathbb{E}[{\tau_{tr}}] + \mathbb{E}[\tau_{ntr}]= \frac{1}{q(1-q)} \label{dummy3}
\end{align}
Thus, inserting (\ref{dummy2}) and (\ref{dummy3}) into (\ref{dummy1}), we obtain 
\begin{align}
\lim_{T\to \infty}\mathbb{E}\left[\frac{\sum_{i=1}^{T}\Delta_{i}(t)}{T}\right] = \frac{1}{q} = N
\end{align}
This is true for all $(N-1)$ non-blocked users. Thus, 
\begin{align}
    \frac{1}{N}\sum_{i=1}^{N}\Delta_{i} \leq \frac{T+1}{2N} + (N-1)
\end{align}
completing the proof.
\end{Proof}

\section{Modified System Model: Sub-Carriers}
\begin{figure}[t]
    \centerline{\includegraphics[width = 0.72\columnwidth]{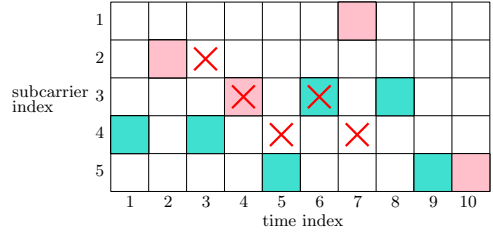}}
    \vspace*{-0.2cm}
    \caption{A pictorial illustration of the modified system model: Pink and turquoise colors represent two users. In each time slot, the BS chooses one user and one sub-carrier to send an update. For example, at time slot $5$, the BS chooses turquoise user, and the $5$th sub-carrier to send an update; and the adversary chooses to block the $4$th sub-carrier.}
    \label{fig8}
    \vspace*{-0.5cm}
\end{figure}

We saw in the previous section that the age increases with $T$ when the BS has only one (dedicated) communication channel to send updates to each user. Here, we propose a modified system model where the BS has $N_{sub}$ sub-carriers to send updates in. The BS can choose any one of these $N_{sub}$ sub-carriers for transmission. We still consider that, at a given time slot, the BS can transmit to at most one user. Thus, at a given time slot, the BS chooses one user and one sub-carrier to send an update. Similar to the previous system model, the adversary aims to block the communication channel so that the age is maximized, with the constraint that, the adversary can block at most $\alpha T$ communication slots; and at a given time slot, the adversary can block at most $1$ communication channel out of $N_{sub}$ communication channels. This modified system model is shown in Fig.~\ref{fig8} for $N=2$ users and $N_{sub}=5$ sub-carriers. 

The adversary can block up to $\alpha T$ time slots. However, the adversary now blocks sub-carriers instead of the users individually. Thus, if $\sigma_{j}(t)=0$, the adversary blocks the $j$th sub-carrier, and if $\sigma_{j}(t)=1$, the adversary does not block the $j$th sub-carrier. Similar to the previous system model, we define a blocking sequence matrix $\sigma$ of size $N_{sub} \times T$ whose $(j,t)$ element is $\sigma_{j}(t)$. Thus, the optimization problem is 
\begin{align}\label{eq:prob_formu2}
      \Delta^{*} = \sup_{\sigma} \inf_{\pi} \quad & \Delta^{\pi,\sigma} \nonumber \\  
       \textrm{s.t.}  \quad & \sum_{k=1}^{N_{sub}} \sum_{j=1}^{T} (1 -\sigma_k(j)) \leq \alpha T\nonumber\\ 
      & \sum_{k=1}^{N_{sub}} (1 - \sigma_k(j)) =1, \quad j=1,\ldots,T
\end{align}

We consider the following scheduling algorithm. At any given time slot, the BS randomly chooses one of the $N$ users with probability  $\frac{1}{N}$, and then randomly chooses one of the $N_{sub}$ sub-carriers with probability $\frac{1}{N_{sub}}$. The BS serves the randomly chosen user in the randomly chosen sub-carrier. 

\begin{theorem}\label{th:3}
For the above mentioned scheduling algorithm, the optimal blocking sequence for the adversary is to block $\alpha T$ consecutive time slots of any particular sub-carrier.
\end{theorem}

\begin{Proof}
We denote $\sum_{k=1}^{N_{sub}} \sigma_{k}(t)$ as $\tilde{\sigma}(t)$. If the adversary does not block any sub-carrier at time-slot $t$, then the age of the $i$th user at time slot $t+1$ will be $(a_{i}(t)+1)$ with probability $\frac{N-1}{N}$ and $1$ with probability $\frac{1}{N}$. If the adversary blocks any one of the $N_{sub}$ sub-carriers at time slot $t$, then the age of the $i$th user at time slot $t+1$ will be $a_{i}(t)+1$ with probability $\frac{N-1}{N} + \frac{1}{N N_{sub}}$ and $1$ with probability $\frac{1}{N} \frac{N_{sub}-1}{N_{sub}}$. Note that, if the adversary blocks any one of the $N_{sub}$ sub-carriers at time slot $t$, then $\tilde{\sigma}(t) = N_{sub} - 1$, otherwise $\tilde{\sigma}(t)=N_{sub}$. Thus, similar to the previous system model, for the $i$th user, 
\begin{align}\label{eq:basic2}
 \Delta_{i}(t + 1) = & \frac{N-1}{N} \left(\Delta_{i}(t)+1\right) + \frac{1}{N}\left(1 - \frac{N_{sub}- \tilde{\sigma}(t)}{N_{sub}}\right)   \nonumber\\ 
 & +\frac{1}{N} \left(\Delta_{i}(t)+1\right) \left(\frac{N_{sub}-\tilde{\sigma}(t)}{N_{sub}}\right)  
\end{align}
Rearranging the terms of (\ref{eq:basic2}), we get
\begin{align}\label{eq:28}
    \Delta_{i}(t+1) = \Delta_i(t) \left(1 - \frac{\tilde{\sigma}(t)}{N N_{sub}}\right) + 1
\end{align}
Note that this is multi sub-carrier version of recursion in (\ref{eq:9}).

As $\tilde{\sigma}(t)$ can take two values, either $N_{sub}$ or $N_{sub}-1$, at time slot $t$, it does not matter which sub-carrier the adversary blocks, it can block any one of the $N_{sub}$ sub-carriers, and still achieve the same $\Delta_{i}(t+1)$. Thus, we assume that the adversary always blocks one particular sub-carrier. Finally, we have to show that the adversary blocks that particular sub-carrier consecutively for $\alpha T$ time slots. Writing (\ref{eq:28}) recursively, 
\begin{align}\label{eq:29}
 \Delta_i(t+1)=&\Delta_i(1) \prod_{j=1}^t \left(1-\frac{{\tilde{\sigma}}(j)}{N N_{sub}}\right) 
 \nonumber\\
 &+ \sum_{\ell=2}^{t} \prod_{j=\ell}^t \left(1-\frac{{\tilde{\sigma}}(j)}{N N_{sub}}\right)+1
\end{align}
which is the multi sub-carrier counterpart of (\ref{eq:11}). Noting that $\Delta_{i}(1)=1$ for all $i=1,\ldots,N$, we obtain 
\begin{align}\label{eq:30}
    \sum_{i=1}^{N}\Delta_i(t+1)= N\left(\sum_{\ell=1}^{t} \prod_{j=\ell}^t \left(1-\frac{{\tilde{\sigma}}(j)}{N N_{sub}}\right)+1\right)
\end{align}

Similar to the previous section, we introduce the concept of trains for ($\ref{eq:30}$). The only difference between the trains of (\ref{eq:11}) and the trains of (\ref{eq:30}) is that the elements of the trains of (\ref{eq:11}) are either $1$ or $\frac{N-1}{N}$, whereas the elements of the trains of (\ref{eq:30}) are either $\frac{N-1}{N}$ or $(1- \frac{N_{sub}-1}{N N_{sub}})$. With similar arguments to Lemma~\ref{lemma:3}, Lemma~\ref{lemma:4} and Lemma~\ref{lemma:5}, we show that blocking consecutive time slots is optimal for the adversary. 
\end{Proof}

In the next theorem, we provide an upper bound for the average age with the proposed algorithm.

\begin{theorem}\label{th:4}
The average age of the above mentioned scheduling algorithm is upper bounded by $\frac{N N_{sub}}{N_{sub}-1}$.
\end{theorem}

\begin{Proof}
The time horizon $T$ can be divided into intervals for a user. The intervals consist of consecutive transmitted time slots and consecutive non-transmitted time slots as discussed in the proof of Theorem~\ref{th:2}. From Theorem~\ref{th:3}, we know that the optimal blocking sequence for the adversary is to block a particular sub-carrier for consecutive $\alpha T$ time slots. Without loss of generality, we assume that the adversary blocks the first sub-carrier. Similar to the proof of Theorem~\ref{th:2}, let us assume that the length of the consecutive transmitted time slots is $\tau_{tr}$ and the length of the consecutive non-transmitted time slots is $\tau_{ntr}$. Then, $\tau_{tr}$ and $\tau_{ntr}$ are geometrically distributed as: $\mathbb{P}(\tau_{tr}=k) = q^{k-1}(1-q)$, and $\mathbb{P}(\tau_{ntr}=k)= q(1-q)^{k-1}$, for $k\geq 1$, where $q=\frac{1}{N}\frac{N_{sub}-1}{N_{sub}}$. Using renewal reward theorem,
\begin{align}
\lim_{T\to \infty}\mathbb{E}\left[\frac{\sum_{t=1}^{T}\Delta_{i}(t)}{T}\right] =\frac{1}{q}= \frac{N N_{sub}}{N_{sub}-1}
\end{align}
Thus, the average age is upper bounded as
\begin{align}\label{eq:upb}
    \Delta \leq \frac{N N_{sub}}{N_{sub}-1}
\end{align}
concluding the proof.
\end{Proof}

\section{Discussion}
From Lemma~\ref{lemma:2}, we know that, for the first system model, the fundamental lower bound is $\frac{T \alpha^{2}}{2 N}$, and from Theorem~\ref{th:2}, we know that for the discussed scheduling algorithm the upper bound is $\frac{T+1}{2N} + (N-1)$. Thus, if $\frac{T}{2 N}\gg \frac{1} {2 N} + (N-1)$, which happens when the time horizon $T$ is large enough, we have
\begin{align}
\frac{\frac{T+1}{2N} + (N-1)}{\frac{T \alpha^{2}}{2 N}} \approx
\frac{\frac{T}{2N}}{\frac{T \alpha^{2}}{2 N}} \approx \frac{1}{\alpha^{2}}
\end{align}
That is, the discussed scheduling algorithm is $\frac{1}{\alpha^{2}}$ optimal. For instance, if the adversary can jam up to half of the time slots, i.e., if $\alpha=\frac{1}{2}$, then the proposed algorithm is $4$ optimal.

From Theorem~\ref{th:4}, we know that the upper bound for the discussed scheduling algorithm for the modified system model is $\frac{N N_{sub}}{N_{sub}-1}$. From \cite[Theorem~1]{banerjee2020fundamental}, the fundamental lower bound for the modified system model can be obtained by assuming that the adversary does not block any of the communication channel, i.e., the probability of successful transmission $p_{i}=1$,
\begin{align} \label{lb_expr}
 \Delta^* \geq 	\frac{1}{2N}	\bigg(\sum_{i=1}^{N} \sqrt{\frac{1}{p_i}}\bigg)^2+ \frac{1}{2} = \frac{N}{2} + \frac{1}{2}
\end{align}
Thus, if $N\gg 1$, i.e., the number of users is large enough, 
\begin{align}
\frac{\frac{N N_{sub}}{N_{sub}-1}}{\frac{N}{2} + \frac{1}{2}} \approx
\frac{\frac{N N_{sub}}{N_{sub}-1}}{\frac{N}{2}} \approx \frac{2 N_{sub}}{N_{sub}-1}
\end{align}
That is, the discussed scheduling algorithm is $\frac{2 N_{sub}}{N_{sub}-1}$ optimal. Thus, the performance of this algorithm increases with the number of sub-carriers. In the worst case scenario the algorithm is $4$ optimal (when $N_{sub}=2$) and it is $2$ optimal when $N_{sub}$ is sufficiently large so that $\frac{N_{sub}-1}{N_{sub}}\approx 1$.

\section{Conclusion}
In this work, we analyzed the average age in the presence of a power constrained adversary. We considered two system models. In the first model, there is no diversity and the adversary can block any one user out of $N$ users at a given time slot. For this case, we saw that the average age increases linearly with $T$. In the second model, we introduced diversity in the form of sub-carriers. For this system model, we assumed that the adversary can block any one sub-carrier out of $N_{sub}$ sub-carriers at a given time slot instead of blocking a user. We observed that, in the presence of diversity the average age is independent of $T$. We also saw that for the modified system model the proposed algorithm is $\frac{2 N_{sub}}{N_{sub}-1}$ optimal.  

\bibliographystyle{unsrt} 
\bibliography{references}
\end{document}